\documentclass[]{interact}

\usepackage{epstopdf}% To incorporate .eps illustrations using PDFLaTeX, etc.
\usepackage[caption=false]{subfig}% Support for small, `sub' figures and tables
\usepackage{hyperref}
\hypersetup{hidelinks,
	colorlinks=true,
	allcolors=black,
	pdfstartview=Fit,
	breaklinks=true}

\usepackage[longnamesfirst,sort]{natbib}% Citation support using natbib.sty
\bibpunct[, ]{(}{)}{;}{a}{,}{,}% Citation support using natbib.sty
\renewcommand\bibfont{\fontsize{10}{12}\selectfont}% To set the list of references in 10 point font using natbib.sty

\theoremstyle{plain}% Theorem-like structures provided by amsthm.sty

\theoremstyle{definition}

\theoremstyle{remark}

\begin{document}
%\bibliographystyle{plainnat}
%\articletype{ARTICLE TEMPLATE}% Specify the article type or omit as appropriate
%\noindent{\large{\textbf{Massive Trajectory Matching and Construction from Aerial Videos based on Frame-by-Frame Vehicle Detections \\}}}

%\noindent{\textbf{Ruyi Feng, Ph.D. Candidate}}

%\noindent{School of Transportation, Southeast University}

%\noindent{2 Si Pai Lou, Nanjing, 210096, China}

%\noindent{Email: fengruyi@seu.edu.cn}

%\\ \hspace*{\fill} \\
%\noindent{\textbf{Zhibin Li, Ph.D., Professor (Corresponding Author)}}

%\noindent{School of Transportation, Southeast University}

%\noindent{2 Si Pai Lou, Nanjing, 210096, China}

%\noindent{Email: lizhibin@seu.edu.cn}

%\\ \hspace*{\fill} \\
%\noindent{\textbf{Changyan Fan, Graduate Research Assistant}}

%\noindent{School of Transportation, Southeast University}

%\noindent{2 Si Pai Lou, Nanjing, 210096, China}

%\noindent{Email: fchyan@seu.edu.cn}

%\clearpage

\title{Massive Trajectory Matching and Construction from Aerial Videos based on Frame-by-Frame Vehicle Detections}

\author{
\name{Ruyi Feng\textsuperscript{a}\thanks{CONTACT Zhibin Li Email: lizhibin@seu.edu.cn}, Zhibin Li\textsuperscript{a} and Changyan Fan\textsuperscript{a}}
\affil{\textsuperscript{a}School of Transportation, Southeast University, Nanjing, China}
}

\maketitle

\begin{abstract}
Vehicle trajectory data provides critical information for traffic flow modeling and analysis. Unmanned aerial vehicles (UAV) is an emerging technology for traffic data collection because of its flexibility and diversity on spatial and temporal coverage. Vehicle trajectories are constructed from frame-by-frame detections. The increase of vehicle counts makes multiple-target matching more challenging. Errors are caused by pixel jitter, vehicle shadows, road marks as well as some missing detections. This research proposes a novel framework for construction of massive vehicle trajectories from aerial videos by matching vehicle detections based on traffic flow dynamic features. The You Look Only Once (YOLO) v4 is used for vehicle detection in UAV videos based on Convolution Neural Network (CNN). Trajectory construction is proposed in detected bounding boxes with trajectory identification, integrity enhancement, and coordinate transformation from image coordinates to the Frenet coordinates. The raw trajectory obtained is then denoised by the ensemble empirical mode decomposition (EEMD). Our framework is tested on two aerial videos taken by a UAV on city expressway covering congested and free-flow traffic conditions. The results show that the proposed framework achieves a Recall of 93.00\% and 86.69\%, and a Precision of 98.86\% and 98.83\% for vehicle trajectories in the free-flow and congested traffic conditions.The trajectory processing speed is about 30s per track.
\end{abstract}


\section{Introduction}
Vehicle trajectories provide vital data support for traffic flow studies. It contains both free-flow and congested traffic states. From the trajectory map, we can calculate macroscopic traffic flow parameters such as average speed, density, and volume, which are the outputs of the widely used inductive loop detectors. The trajectory map also provides an intuitive view of traffic phenomena such as kinematic wave and traffic breakdown. Moreover, it includes traffic parameters at the microscopic scope such as speed; time headway and space headway of individual vehicles. Previously, numerous studies have utilized vehicle trajectory data for traffic flow model calibration ( e.g. \citealp{TO2010, Zheng:2013,Talebpour:2015}), traffic feature/phenomena exploration, (e.g. \citealp{Chen:2012,Coifman:2015}), car-following and lane-changing behavior analysis, (e.g. \citealp{Siqueira:2006,Herrera:2010}), and driving strategy development, (e.g. \citealp{Li:2012,Rhoades:2016,Aghabayks:2013}).

The most famous vehicle trajectory dataset is the Next Generation Simulation (NGSIM) launched by the Federal Highway Administration. The project installs multiple fixed cameras on the top of a nearby building to collect traffic videos. It provides vehicle trajectory data of four road segments, containing information such as instantaneous vehicle velocity, acceleration, position coordinates, vehicle length, and vehicle type. This dataset has been widely used since its release \citep{He:2017}. However, the dataset has some limitations such as fixed road segment, insufficient coverage range, limited traffic flow condition, limited vehicle component, as well as erroneous speed and acceleration information \citep{Montanino:2013}. Such limitations are associated with the fixed locations of cameras at inclined shooting angles as well as the trajectory extraction methods applied on those videos (e.g. \citealt{Montanino:2013,Punzo:2011,Zheng:2012}). A multi-step reconstruction algorithm was provided for verifying abnormal trajectories in this dataset, e.g.\citep{Montanino:2015}.

In recent years, unmanned aerial vehicles (UAV) technology has emerged in traffic data collection because of its high flexibility, low cost and broad covering range. Previous studies have proposed approaches to extract vehicle trajectories from UAV videos. The critical challenges include small-sized objects, indistinguishable features, the influence of shadows and simultaneous operation for multiple vehicles. e.g.\citep{Emmanouil:2016,Feng:2020}. Some studies apply feature-based tracking algorithms to track vehicles in consecutive frames based on the maximum similarity criterion. They usually need the initial positions of the to-be-tracked targets in the region of interest (ROI) which locates in upstream. For example, \citet{Xiao:2012} apply a Sniff object-tracking algorithm with the optical flow for vehicle position initialization. \citet{Saleemi:2013} propose a multi-frame many-many point correspondence-tracking algorithm using the Harris corner detector. \citet{Chen:2019} combines different feature descriptors to form a multi-view learning and sparse representation-based target tracker. However, the performance of the tracking-based trajectory extraction algorithms is very sensitive to the initialization of vehicle position. In other words, the trajectory will be entirely lost if it is not detected in the ROI.  

To avoid such issue, studies construct the vehicle trajectories by detecting vehicles in the entire spatial-temporal region of the videos. For example, \citet{Cao:2011} use a support vector machine (SVM) classifier with the bLPS-HOG features for vehicle detections. They then applied a spatiotemporal appearance-related similarity method (STARS) for target correlation. \citet{Apeltauer:2015} use the AdaBoost classifier to learn the multi-scale block local binary patterns features for detection and tracked vehicles with the single-particle filter. \citet{Azevedo:2014} adopt the background subtraction for vehicle detection, and the k-shortest disjoints paths algorithm for data correlation. The performance is highly affected by the vehicle detection accuracy. Specifically, if the loss rate of detections exceeds some thresholds, vehicles could be mismatched in different frames and erroneous trajectories are produced. Thus, how to accurately match vehicles from massive frame-by-frame detection records to construct true trajectories is rather challenging.

This research aims at proposing an accurate vehicle matching and trajectory reconstruction framework using frame-by-frame vehicle detection data. The traffic dynamic features in massive vehicle flow are applied to generate constraints for trajectories matching and construction. The performance of our proposed method is tested on two aerial videos with different target quantities and traffic conditions. We also estimate the processing speed of our trajectory construction method. 

\section{METHODS}

\subsection{Overall Framework}

\begin{figure}
	\centering
	\includegraphics[width=0.9\linewidth]{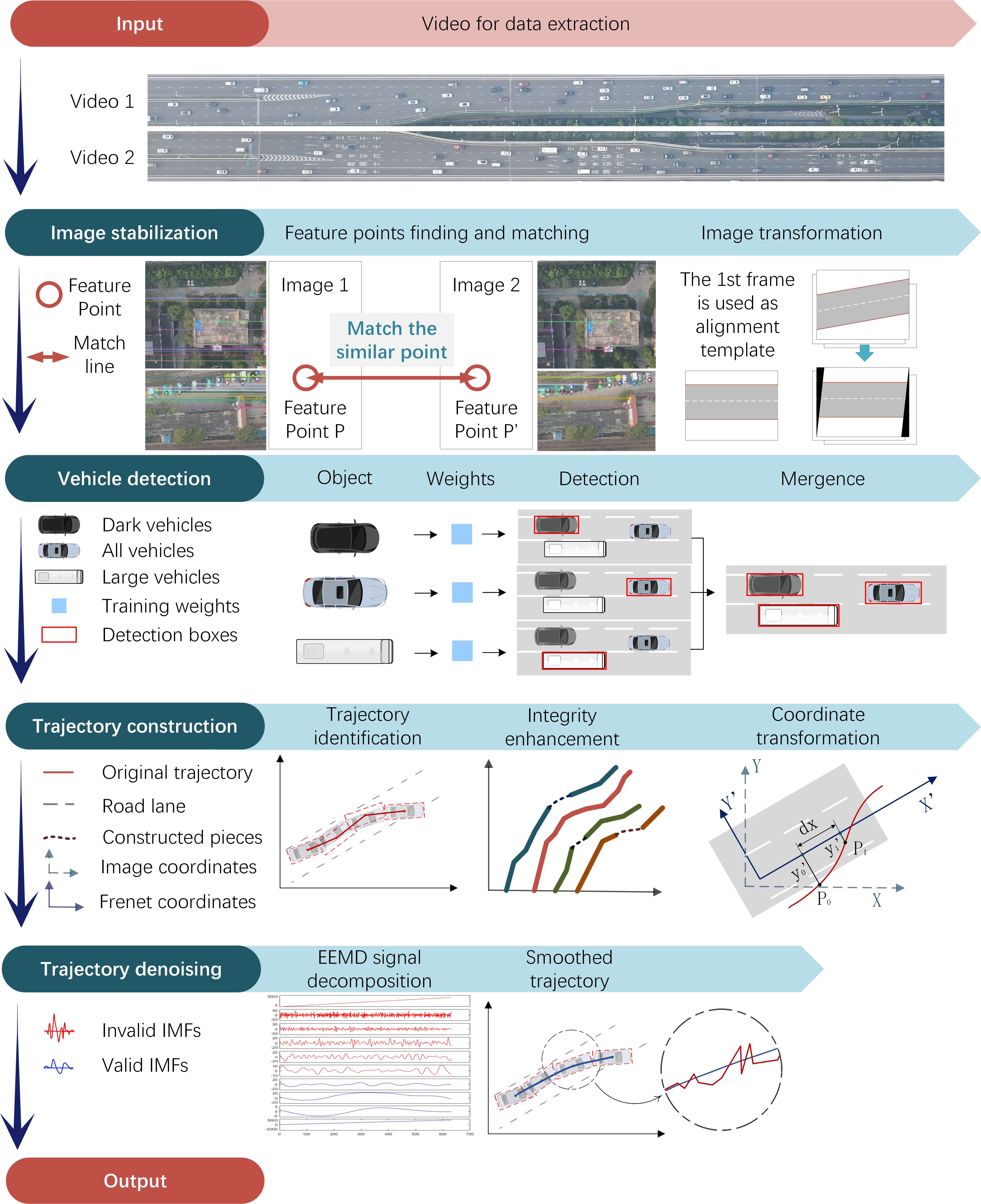}
	\caption{Framework flow chart of our algorithm.}
	\label{fig:1}
\end{figure}

Our framework contains four modules, as shown in Figure~\ref{fig:1}, which are image stabilization, vehicle detection, trajectory construction, and trajectory denoising. In the first module, shaking in UAV videos is eliminated by Scale Invariant Feature Transform (SIFT) operator matching. YOLOv4 algorithm for vehicle detection is applied for position obtaining as the foundation of trajectory construction. Then a data correlation algorithm based on traffic flow theory combined with regional traffic dynamics is proposed to construct vehicle trajectories. In the end, noise in rough vehicle trajectories is separated and removed by the EEMD-based denoising algorithm. The construction details of each section are described below.

\subsection{Image Stabilization}

UAV may shake due to the interference of airflow and wind (e.g.\citealp{Li:2019}).The external disturbances could cause viewing angle changes in the image. Such movement will overlap with the target movement in the video and cause an error. In image stabilization, a feature point selection method using the scale-invariant feature as decision method, SIFT \citep{Lowe:2004} is applied. SIFT is an image local feature description operator based on scale-space that maintains invariance to image scaling, rotation, and even affine transformation. The key points that SIFT looks for are some very prominent "stable" feature points that will not change due to factors such as illumination, affine transformation and noise, such as corner points, edge points, bright spots in dark areas, and dark points in bright areas. 

The selection of feature point pairs uses the joint restriction of Euclidean distance and location range. The Euclidean distance is to ensure the similarity of the two feature points to the internal scale-invariant vector. And the location range is to exclude the mismatch caused by different objects had similar feature vectors based on no violent motion of UAV. Based on these two constraints, all video frames will be affine transformed by the good-key-point pairs found by the SIFT operator and aligned with the first frame. For stronger fault tolerance, the multiple good-key-point pairs found will all be used for matching operations. The calculated affine transformation matrix searched by the minimum sum of errors for all good-key-point.

\subsection{CNN-based Vehicle Detection}
YOLO series (\citealp{Redmon:2016}; \citealp{Bochkovskiy:2020}) and R-CNN \citep{Fan:2016} series are two well-known masterpieces  of detection in the field of CNN (Benjdira et al., 2019).Some scholars have conducted an experimental comparison between respective outstanding works of the two series, YOLOv3 and Faster-RCNN. This research shows more suitability for vehicle detection in aerial videos of YOLOv3 and achieves higher detection accuracy and faster processing speed than R-CNN \citep{Benjdira:2019}. The current YOLO has been upgraded to version 4 \citep{Bochkovskiy:2020}, which fully integrates computing skills and has a significant improvement in speed and edge accuracy. Note that the latest updated version, YOLOv5 \citep{jocher2020yolov5}, emphasizes its lightweight and training speed, and its accuracy can almost reach the same as YOLOv4. As the detection accuracy is concerned most, YOLOv4 is used in this research.  

\begin{figure}
	\centering
	\includegraphics[width=0.7\linewidth]{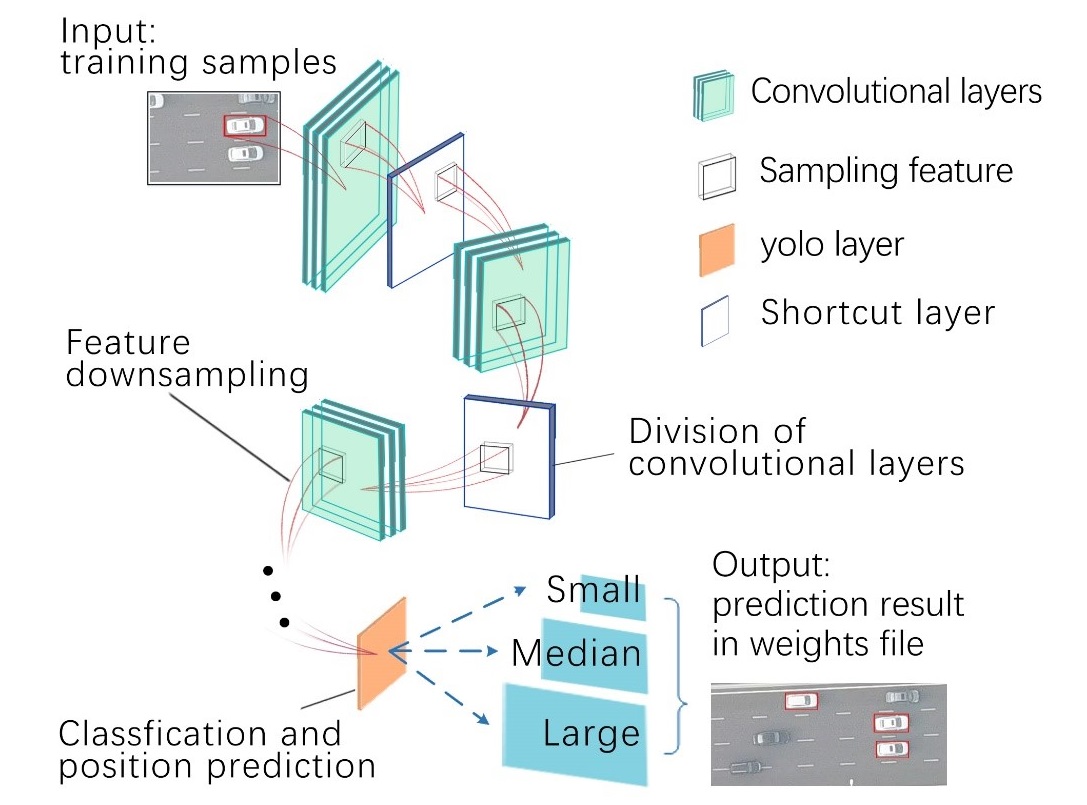}
	\caption{YOLOv4 natural work schematic diagram.}
	\label{fig:2}
\end{figure}

In YOLOv4, the convolutional layer and sampling layer segment operate the image alternating continuously (Figure~\ref{fig:2}). The input image first goes to convolutional layers to get samples of features. And then, 1×1 or 3×3 layers resetting is applied for quicker calculating speed. The shortcut layers are arranged among convolutional layers at specific intervals to divide the CNN into dozens of pieces. The dividing operation is necessary for controlling the propagation of gradients as well as avoiding the problems of gradient diffusion and gradient exploding. At last, the yolo layers perform classification and position prediction of the targets. The entire calculation process and target characteristics will be continuously summarized and updated as a weights file. This weights file is the database of target detection.

% TODO: \usepackage{graphicx} required

\begin{figure}
	\centering
	\includegraphics[width=0.9\linewidth]{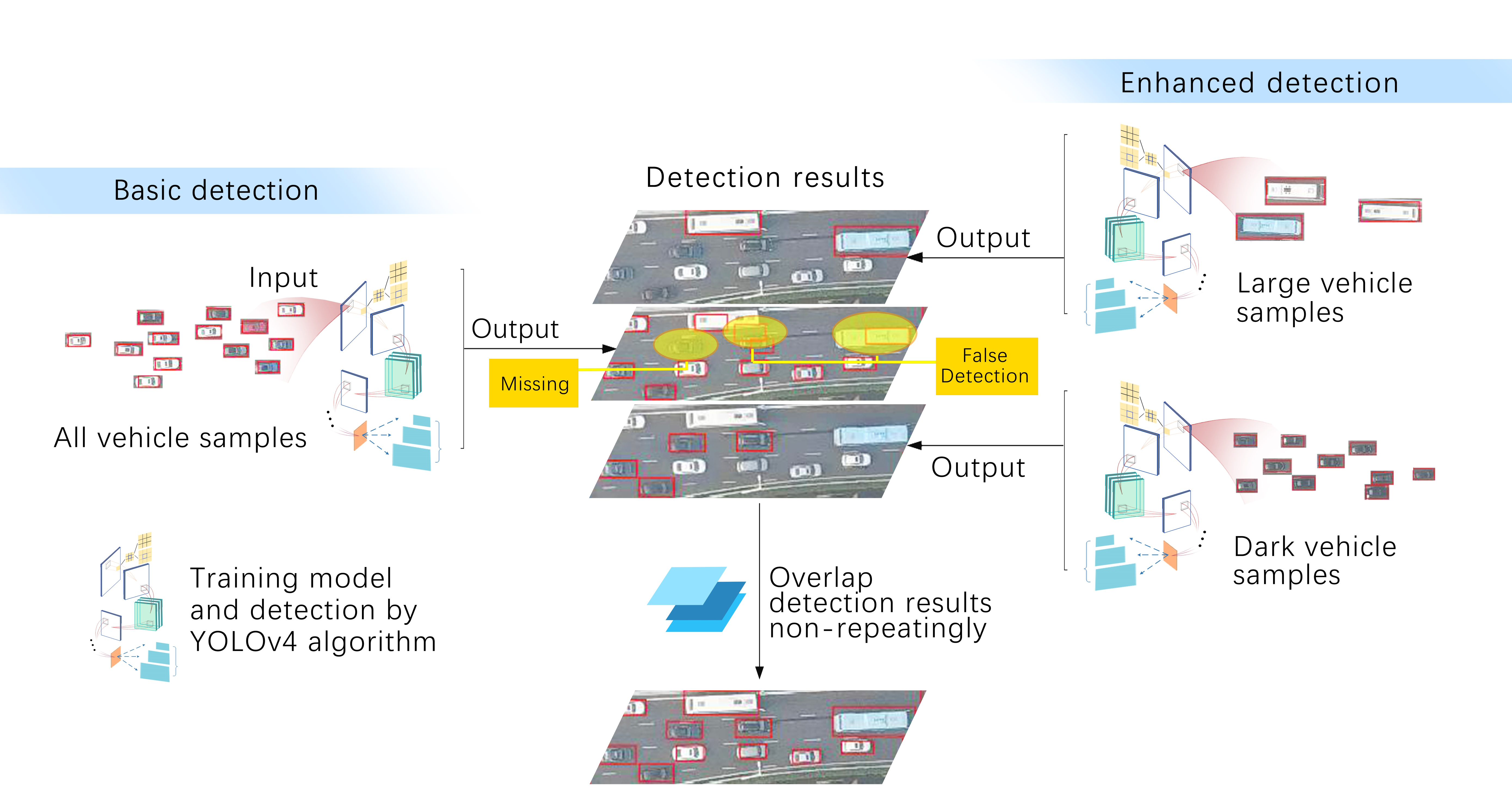}
	\caption{The merged basic and enhanced training strategy.}
	\label{fig:3}
\end{figure}

The model performs instable in dark vehicle and large vehicle detection because of their distinguished features and small training samples. To solve this problem, we propose a merged training strategy to improve detection performance (see Figure~\ref{fig:3}). We first train the YOLOv4 model with all vehicle samples in the training database. In the second step, we further enhance our method by separately training two additional neural network models for dark vehicles and large vehicles. After that, the vehicle detections from the three models are merged to form the raw detection pool.

% TODO: \usepackage{graphicx} required

The raw data may contain duplicated detection boxes of the same vehicle, which need to be reduced. Intersection-over-union (IoU) is used in the duplication reduction method, the calculation of IoU is shown in Figure~\ref{fig:4},  and the recognition mechanism is expressed as (Eq.~\ref{eq:1},~\ref{eq:2},\ref{eq:3}):
% TODO: \usepackage{graphicx} required

\begin{figure}
	\centering
	\includegraphics[width=0.6\linewidth]{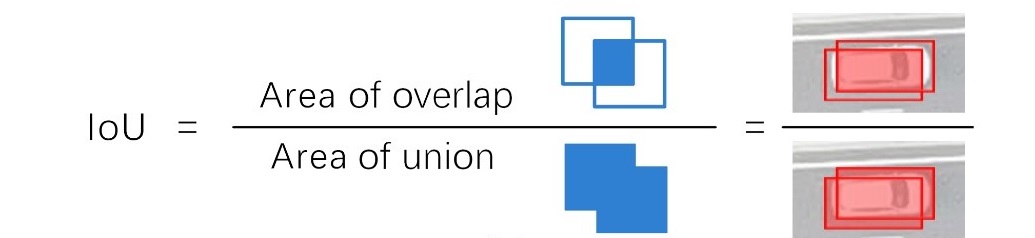}
	\caption{The calculation of IoU.}
	\label{fig:4}
\end{figure}

\begin{equation} \label{eq:1}
	max(IoU_{obj1}, IoU_{obj2})\leq area_{overlap} \;
\end{equation}
\begin{equation} \label{eq:2}
	area_{overlap} \leq min ( IoU_{obj_1}, IoU_{obj_2}) \;
\end{equation}
\begin{equation} \label{eq:3}
	IoU_{obj1}\leq area_{overlap}<IoU_{obj2} \;
\end{equation}
where $ IoU_{obj1},IoU_{obj2} $ indicate the IoU value of two overlapped object obj1 and obj2, $ area_{overlap} $ is the maximum acceptable percentage of target bounding box overlapping. In Eq.~\ref{eq:1}-\ref{eq:3}, only one item will be established for each judgment. When it suits Eq.~\ref{eq:1} or Eq.~\ref{eq:2}, the higher confidence object will be reserved.And if Eq.~\ref{eq:3} holds, the lower overlapped obj1 will be choose.

After completing the repeated screening, the location of the vehicle at each moment is obtained. At this time, we only have a scattered distribution of points without any links. In the trajectory construction part, these points will be connected in series to construct the trajectory.

\subsection{Trajectory Construction}
A large number of vehicle targets are contained in every frame. Errors are also contained which could be caused by pixel jitter, vehicle shadows, road marks as well as some missing detections. Such errors make it more difficult to matching construction in massive vehicle candidates. Directly linking these detection frames with errors result in irreparable effects on traffic data. Existing association methods are processed in a mixture of these right and wrong candidate boxes and result in a high error rate (\citealp{Azevedo:2014};  \citealp{luetteke2012implementation}).

It is crucial to choose the only true position of one vehicle among the candidates mixed with interference target. And in an individual vehicle, the position during travel space must be determined by vehicle dynamics and surroundings. In other words, by combining vehicle dynamics and surroundings, we can predict a more reliable position in multiply candidates. In this section, traffic flow theory is used to deduce the correct way to match vehicles. In addition, under the constraints of the time section, there will still be some unreasonable fragments in the trajectory. The unreasonable fragments should be cut, and the vacant parts should be compensated. In the following section, trajectory identification integrity enhancement and coordinate transformation are introduced.

\subsubsection{Step 1: Trajectory Identification}

At the beginning of this step, it is necessary to apply an explanation of traffic flow theory in differential conditions proposed by \citet{newell1993} The original equation is as Eq.~\ref{eq:4ad}. Then we do a conventional deformation and it looks like Eq.~\ref{eq:5ad}, where $\Delta Q$ is the inflow (or increase) flow in the differential area, $\Delta speed$ is the speed increment in this area and $\Delta k$ indicates the density increment of this area. 

\begin{equation} \label{eq:4ad}
	\frac{\partial k(x,t)}{\partial t}+\frac{\partial q(x,t)}{\partial x}=0 \;
\end{equation}
\begin{equation} \label{eq:5ad}
	\Delta Q=\Delta speed \times (-\Delta k)
\end{equation}

In the proposed explanation, we will redefine Eq.~\ref{eq:5ad} from the perspective of tendency. That is, $\Delta Q$ is defined as the inflow tendency of vehicles, $-\Delta k$ is redefined as the regional gravity. When either the $\Delta speed$ or the $-\Delta k$ becomes larger, the stronger tendency of the vehicle flows into. Besides, we can also redefine regional gravity as Eq.~\ref{eq:6ad}. It shows that the definition of density. For a wider use of this equation, we assume that there are N (N is a constant number) invisible vehicles in every differential region. As a result, in the gap area, the larger of space, the larger of $-\Delta k$. Then, it causes a positive value of regional gravity ($-\Delta k$) and a stronger vehicle inflow tendency.

\begin{equation} \label{eq:6ad}
	-k=-\frac{n}{space}
\end{equation}

Such an inflow tendency explanation can also be confirmed from the car following and lane changing models. Whether in the IDM, Gipps model of car-following(\citealp{treiber2000IDM}; \citealp{gipps1981behavioural}) or the MOBIL model, CA model(\citealp{treiber2016mobil}; \citealp{nagel1998CA}) of lane-changing, the driver will continuously determine the vehicle's running speed and direction according to the gap size during driving.

In the proposed inflow tendency explanation, two critical factors affect the inflow trend. One is regional speed, and the other is the density of the differential area. In this step of trajectory construction, the two factors will be both used as the limitation for an accurate candidate selection.

\begin{figure}
	\centering
	\includegraphics[width=0.8\linewidth]{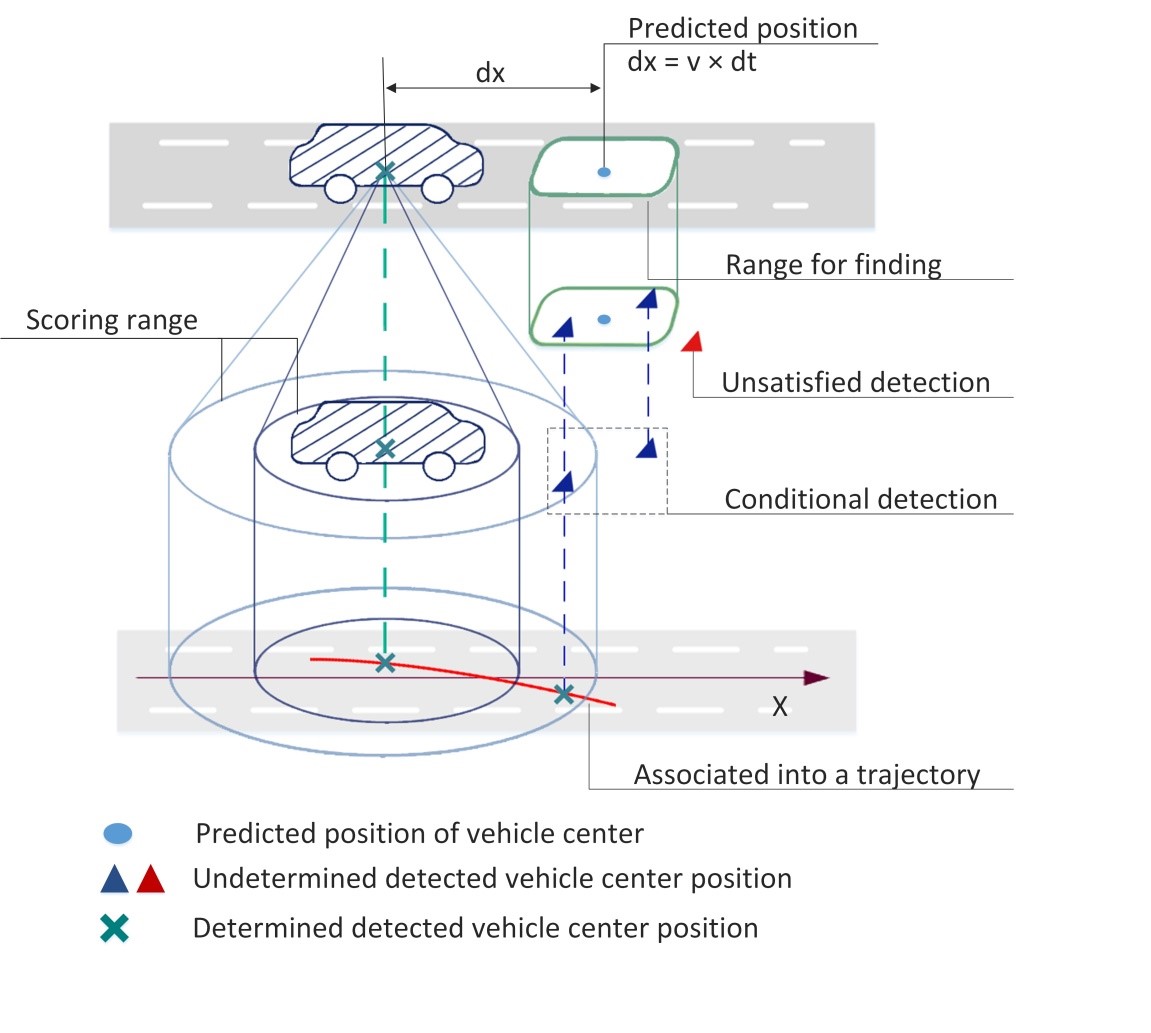}
	\caption{Schematic diagram of identification in speed searching step.}
	\label{fig:5}
\end{figure}

We first use speed as a broad constrain. The vehicle position distribution of the former and the later frames are mainly applied, and the procedure is shown in Fig.~\ref{fig:5}.For the detection boxes in the former frame, we conduct a kinematic prediction and find its possible position in the next frame. In the later frame, the single object of position will be directly categorized as the same vehicle. But sometimes there are multiple choices, maybe a shadow detection box. Thus scoring methods are proposed to determine the most likely one.

% TODO: \usepackage{graphicx} required

\begin{figure}
	\centering
	\includegraphics[width=0.8\linewidth]{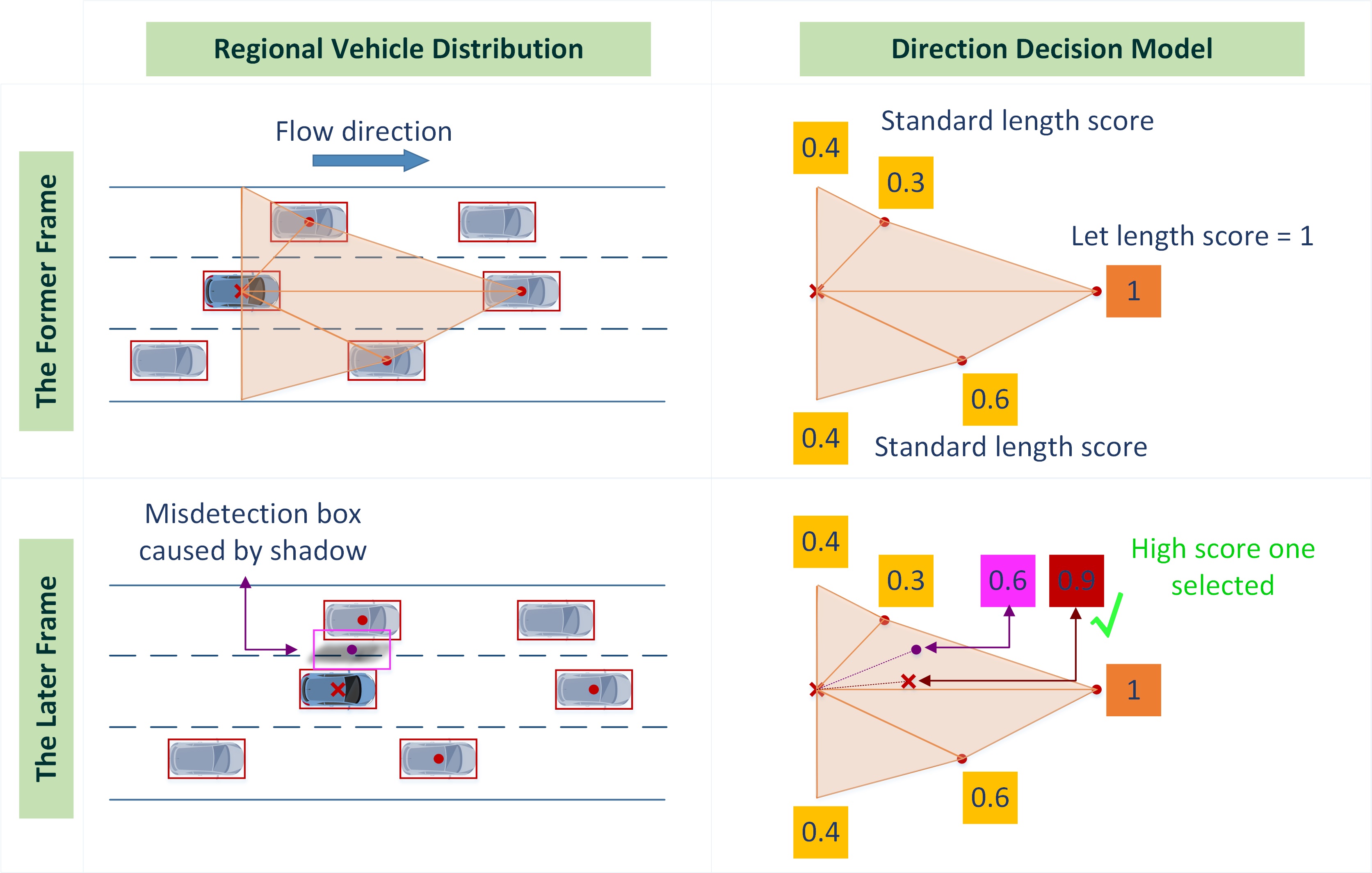}
	\caption{Regional vehicle cooperative direction decision model.}
	\label{fig:6}
\end{figure}

Within a certain speed range, it needs to select the driving direction. In the proposed method, we assume the vehicle will move in the direction of greater clearance. So, the vehicle distribution around the target in the former frame is collected and used to calculate the probability of movement in each direction. As shown in Fig.~\ref{fig:6},  in the direction decision model, the largest gap will be standardized as 1, and the link to other vehicles will be expressed in a ratio. This ratio will be regarded as the potential of each direction, and the candidate box with a high potential in the later frame will be selected. Only the angle is considered, and the candidate frame at the angle between the two inflection points will be scored by interpolation. Due to the different positions and different influences of surrounding vehicles, even if the candidate shadow box is very close to the target vehicle, its direction of arrival should be different from the candidate of the target vehicle.

% TODO: \usepackage{graphicx} required

It needs to notice that once a frame is not detected, the trajectory will end. Therefore, to be better compatible with the identification problem, we will continue to look for the future trajectory of the target vehicle. This search cannot be unlimited, so $TH$ is proposed to restrict the maximum number of future frames in the search process. In our study, we conduct multiple tests and finally set $TH$ to be 5, which has the best performance in our experiments. If a detection box is not detected in the next five frames, it is considered a false one and removed from the trajectory data. If the box can be continuously identified, its position is recorded to form the trajectory track for the vehicle in the following frames. The procedure repeats until there is no detection box to be correlated, and the trajectory is considered ended. The above process is conducted for every detection box in all frames. In this way, the raw trajectory data is produced. 

\subsubsection{Step 2: Integrity Enhancement}

The trajectory obtained in the above steps may have false positives. These false positive parts are called ghost track hereinafter. The following two steps are applied for processing these unreasonable fragments.

1) Ghost track fragments are taken out first. The movement of ghost track fragment does not match the characteristics of driving. It may have an unfixed moving direction and abnormal changes of speed. Non-linear stimulus car-following model (Eq.~\ref{eq:7ad} was applied in each trajectory to check whether the position and speed are valid.$X_{n}(t)$ indicates the position of vehicle n in time $t$, $\dot X_{n}(t)$ is speed and $\ddot X_{n+1}$ is acceleration. The first second of the trajectory is used to initialize the parameter $\lambda_1$ and reaction time $T$. During the verification, the leader vehicle will update as its location changing. When the leader is not found or changes, it will be treated as no stimulation until the vehicle in front is stable. Then the entire trajectory with anomalies and the abnormal trajectory parts at the beginning and end will be removed.

\begin{equation} \label{eq:7ad}
	\ddot X_{n+1}(t+T)=\frac{\lambda_{1}}{[X_{n}(t)-X_{n+1}(t)]^2} [\dot X_{n}(t)-\dot X_{n+1}(t)] \;
\end{equation}

2) Reasonable trajectories will be connected accordingly. After the above processing, the trajectory fragments that come to this step are reasonable but may be broken. These broken pieces will be compensated using quantic polynomial fitting based on the assumption that rate of acceleration change is in a uniform change speed. 

\subsubsection{Step 3: Coordinate Transformation}

The purpose of the coordinate transformation is to convert the vehicle position information from the video image coordinates (where the axes are the border of the video) to the so-called Frenet coordinates (e.g.\citealp{Mao:2015}; \citealp{Malik:2003}), whose axes are along and vertical to the road lane. The details of the coordinate transformation algorithm are presented as follows.

Firstly, road-lane-curve information is needed for establishing axes of Frenet coordinates. Considering the UAV camera is stationary during data collection, we manually collect sample points of lane boundaries and use the polyfit function to fit the lane curves. Each lane center curve is obtained by taking the average of its upper and lower lane boundaries. The UAV may move slightly due to the interference of wind or signal instability. As a result, we perform the lane fitting procedure multiple times and update the coordinates to ensure their accuracy. 

\begin{figure}
	\centering
	\includegraphics[width=0.9\linewidth]{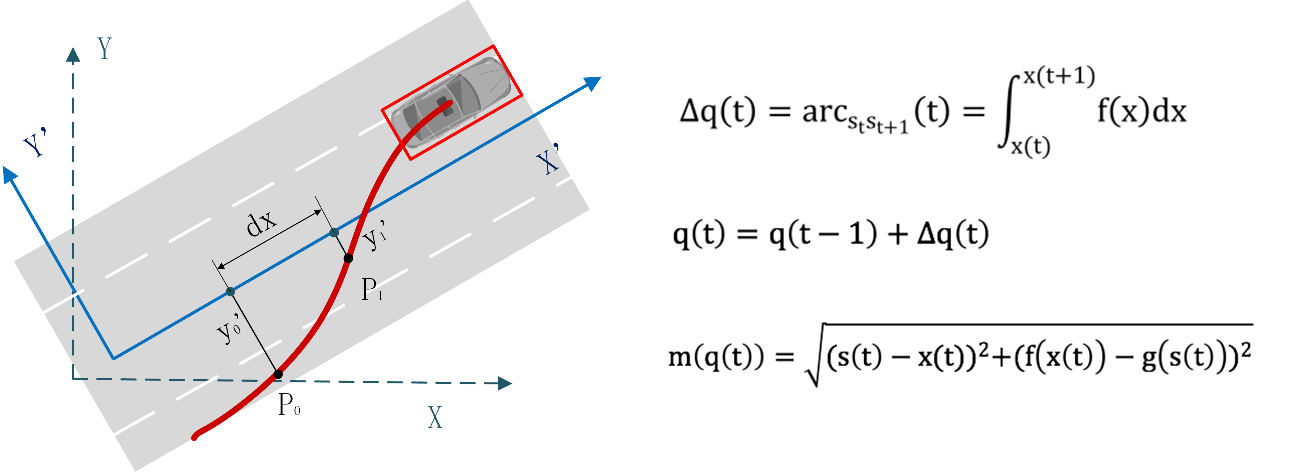}
	\caption{Sketch map of transforming image coordinates to Frenet coordinates. }
	\label{fig:7}
\end{figure}

Since we have known the lane information, the current vehicle position under Frenet coordinates can be calculated as follows (Fig.~\ref{fig:7}). Where $ \{x(t),f(x(t))\},(t=0,1,2...)$ represents vehicle position at time t, and $ \{s(t),g(s(t))\},(t=0,1,2,..) $ is the point closest to the current vehicle position on the lane center curve. $ q(t) $ is the vehicle movement along the road lane, m(q(t)) is the vehicle movement vertical to the road lane. $ \triangle q(t) $ is the movement of q(t) in each frame. Note that if taken each frame as a microelement, $ \triangle q(t) $ can be approximated to the distance between current and last mapping point $ \{s(t), g(s(t))\} $. After the above steps, the trajectory data in the Frenet coordinates is established.

% TODO: \usepackage{graphicx} required

\subsection{Trajectory Denoising}

The trajectories contain noises with features of high frequency and small magnitude due to the limitations of integer pixels and background interference. Although these deviations are composed of white noise, they will still have an impact on the calculation of instantaneous velocity and acceleration. The ensemble empirical mode decomposition (EEMD) is applied for our research \citep{Wu:2009}. EEMD have a good performance for both linear and non-linear signal decomposition. Besides, the decomposed multiple intrinsic mode functions (IMF) in different frequencies can robustly avoid modal mixing, which indicates either an IMF consisting of signals of widely disparate scales, or a signal of a similar scale residing in different IMF components. 

In the trajectory decomposition, in order to better determine the frequency band of each IMF, the additional white noise of finite-amplitude is added to the IMF. Accurately, after adding white noise, the current observation can be described as follows (Eq.~\ref{eq:4}):

\begin{equation} \label{eq:4}
	X_i (t)=x(t)+w_i (t) \;
\end{equation}
where $i$ is the iteration number, $x(t)$ is the raw trajectory data, $w_i (t)$ is the added white noise, $X_i (t)$ is the noise-aided trajectory data.

We decompose the trajectory through the following rules. We first line maximum points and minimum points with cubic splines to form the upper and lower envelopes. Then the local means of the upper and lower envelope are taken as the next observation $x_{i} (t)$. The IMF is found if $x_{i} (t)$ meets the following conditions (Eq.~\ref{eq:5},~\ref{eq:6}):

\begin{equation} \label{eq:5}
	N_z-1\leq N_e\leq N_z+1 \;
\end{equation}
\begin{equation} \label{eq:6}
	\frac{f_{max} (t)+f_{min} (t)}{2}=0 \;
\end{equation}
Define the extrema of $x_i (t)$ as $N_z$, the zero-crossing point as $N_e$, the upper and lower envelope as $f_{max} (t)$ and $f_{min} (t)$. 

When IMFs can no longer be found, this searching will end and original $X_i (t)$ can be expressed as Eq.~\ref{eq:7}:

\begin{equation} \label{eq:7}
	X_i (t)=\sum_{j=1}^{n}imf_j(t)+r_n (t) \;
\end{equation}
where $imf_j (t)$ is the IMFs separated from $X_i (t)$, andv$ r_n (t)$ is the residual.

Appropriate IMFs are determined with an energy-based IMF selection method \cite{Chen:2018}. In this research, it shows that the energy of the signal represents the amount of information stored in it. The energy of signal can be calculated as Eq.~\ref{eq:8}. The energy of noise signal only accounts for a small proportion while the majority of energy is concentrated in the meaningful signals. A log function is used to distinguish the noise signal energy which processes a negative result, and the other meaningful signals need to meet the Eq.~\ref{eq:9}.

\begin{equation}\label{eq:8}
	E_j =\frac{1}{num}\sum_{k=1}^{num}[c_j(k)]^2
\end{equation}
\begin{equation}\label{eq:9}
	log_2E_j>0
\end{equation}
where $E_j$ is the energy of the jth IMF, $c_j (k)$ is the point collection of jth IMF, num is the total number of collection points. The sum of the selected IMFs is the denoised trajectory of the vehicle. After the above steps, accurate vehicle trajectories are extracted via our framework.

\section{EXPERIMENT DESIGN}

The proposed framework is evaluated on two aerial videos captured by high definition camera mounted on a UAV (DJI Mavic professional). Videos are captured at 24 frames-per-second (fps) and with the resolution of 4096 pixel ×2160 pixel. The two test videos were shot at different times of the urban expressway in Nanjing, China. Test video \#1 is taken at a 280m altitude in a free flow traffic condition, while test video \#2 is taken at a 310m altitude in a congested condition. In the course of data acquisition, the UAV is set to be stabilized in the air. The video information is shown in Table~\ref{table:1}.

\begin{table}
	\tbl{Test Video Information}
	{\begin{tabular}{l l l}\toprule
			\multicolumn{1}{c}{Video Information} &
			\multicolumn{1}{c}{Test Video \#1} &
			\multicolumn{1}{c}{Test Video \#2} \\ \midrule
			\multicolumn{1}{c}{Traffic condition} &
			\multicolumn{1}{c}{Free flow} &
			\multicolumn{1}{c}{Congested} \\
			\midrule
			Road geometry & Curve & Curve  \\
			Frame rate & 24fps & 24fps \\
			Resolution & $4096 \times 2160$ & $4096 \times 2160$ \\
			Length & 386 & 427 \\
			Duration & 333s & 300s \\
			Frame & 7992 & 7200 \\
			Focal length & 23mm & 23mm \\
			Flying height & 280m & 310m \\
			\bottomrule
	\end{tabular}}
	\label{table:1}
\end{table}

The experiment is launched on a workstation with a 2.5 GHz E5-2678v3 dual processor and a NVIDIA 2080 Ti graphics card. The calculation and processing are developed on the platform of Visual Studio 2015 and Matlab 2016a.

In terms of model evaluation, we use counting methods to evaluate the completeness of large-scale trajectories. The confusion matrix and two goodness-of-fit measures are quoted to evaluate the accuracy of the trajectory counting results (Table~\ref{table:2} and Eq.~\ref{eq:10}, ~\ref{eq:11}). And also, for the accuracy of trajectory details, the research group manually mark the ground truth and calculate the IoU with the extracted trajectory. 

\begin{table}
	\tbl{Confusion Matrix}
	{\begin{tabular}{l l l}\toprule
		Ground Truth & \multicolumn{2}{c}{Model Results} \\~ & \multicolumn{1}{c}{True} & \multicolumn{1}{c}{False} \\ \midrule
		True & True positive (TP) & False Positive (FP) \\
		False & False Negative (FN) & True Negative (TN)  \\ \bottomrule
	\end{tabular}}
\label{table:2}
\end{table}

\begin{equation}\label{eq:10}
	Recall=\frac{TP}{GT}
\end{equation}
\begin{equation}\label{eq:11}
	Precesion=\frac{TP}{TP+FP}
\end{equation}

\section{RESULTS}

\subsection{Results of Image Stabilization}

All frames of UAV videos will be matched with the first frame. 200 strongest initial feature points are selected and matched in every frame by the SIFT calculator. Not all matched couples are effectively the same point, so the Euclidean distance of the feature invariant vectors of the two matching points is used to test and screen the effective feature point pairs. In order to better obtain effective feature points, we limit the search range of feature points to outside the road area. Most of the images in these areas are static buildings, which can provide effective feature points. 

\begin{figure}
	\centering
	\includegraphics[width=0.7\linewidth]{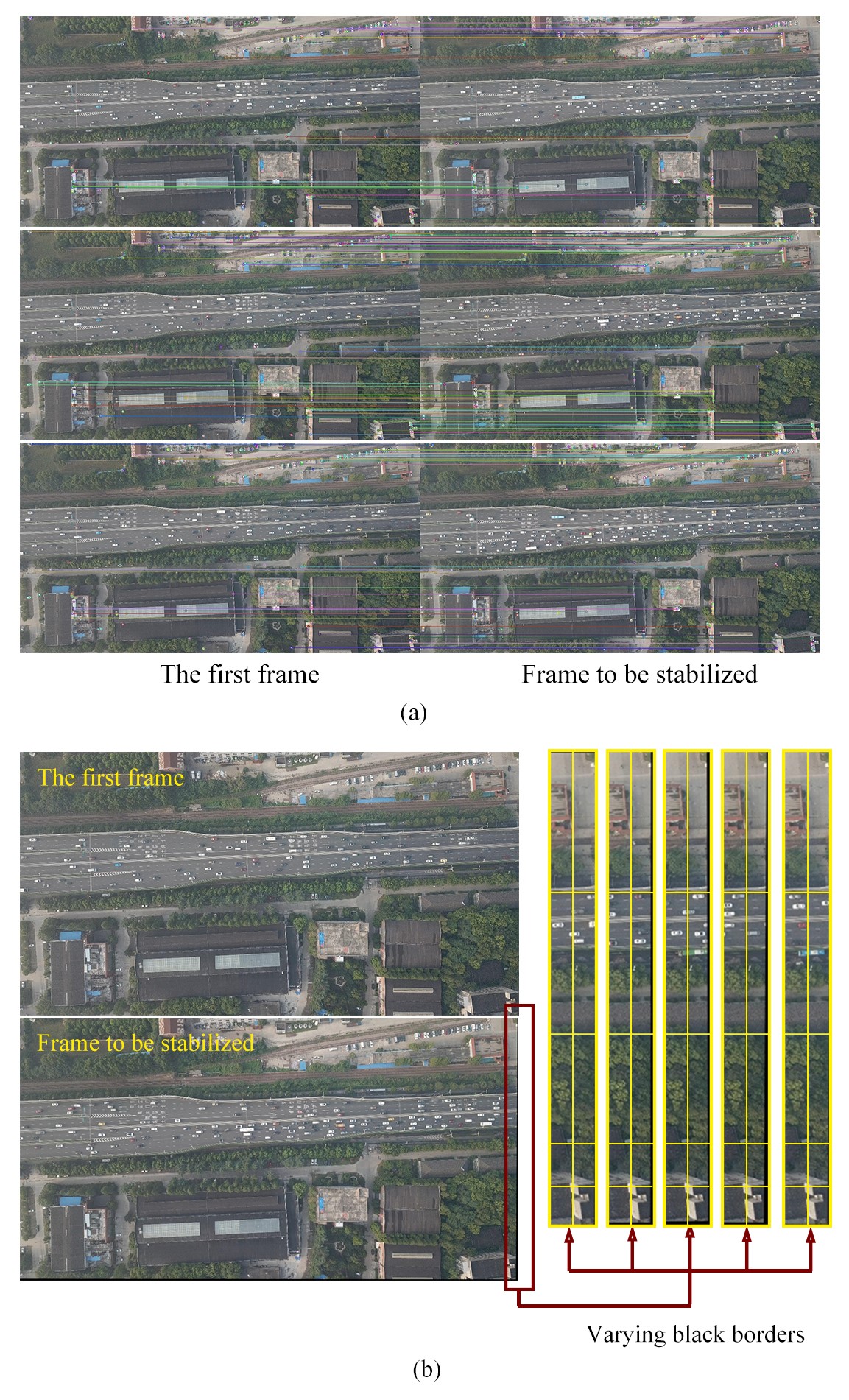}
	\caption{After image stabilization, each pixel matched and the non-pixel portions in borders area are filled with black.}
	\label{fig:8}
\end{figure}

As shown in Figure ~\ref{fig:8}(a), the colored circles represent the acquired feature points, and the colored lines connect the effective feature point pairs. The proportion of retained results after the screening is not large, but they are all exact matches. Registration requires at least three pairs of points to achieve affine transformation. Now, these dozens of matching points have been able to ensure the registration of the picture, which will match based on the principle of minimum error. The stabilized image is shown in  Figure ~\ref{fig:8}(b). It can be seen that the characteristic objects are all in the same position after the image changes, and the non-pixel portions in border areas are filled with black.

% TODO: \usepackage{graphicx} required

\subsection{Model Training and Parameter Calibration}

The basic network and an enhanced network of YOLOv4 are trained for aerial vehicle detection. According to the creator and some researchers who have applied YOLOv4, the training settings and training samples almost play a decisive role in the performance of the network. In our study, 20000 vehicle samples are applied for the basic training. Besides, 4000 dark vehicle samples and 1000 large vehicle samples are applied for enhanced training. These datasets are manually calibrated to ensure that the external edge error is within three pixels. 

Critical parameters of YOLOv4 and the values we choose are listed in Table\ref{table:3}. The Batch indicates the number of samples that participated in each iteration. A higher batch will bring a better accuracy but a lower training speed. The subdivision is set to relieve the usage of GPU by separating training datasets into groups. Width and Height are the resized image specifications for training. Large Width and Height will retain more image information but consume more memory. Max-batches is the upper limit of training times. The best value of the Max-batch is to get less loss, but overlarge Max-batch can easily overfit. The learning rate is to changes the model in each based on training results. A higher learning rate is required at beginning to get a quick decay in loss and low down in the back to ensure learning accuracy and prevent overfitting.

\begin{table}
	\tbl{YOLOv4 Training Parameters}
	{\begin{tabular}{l l l l}\toprule
	Configuration parameters &
	Basic detection &
	\multicolumn{2}{c}{Enhanced detection} \\
	~ & ~ & \multicolumn{1}{c}{Dark vehicle detection} &
	\multicolumn{1}{c}{Large vehicle detection}\\ \midrule
	Number of training samples & 20000 & 4000 & 1000 \\
	Batch & 128 & 128 & 128  \\
	Subdivisions & 32 & 32 & 32 \\
	Width & 672 & 672 & 672 \\
	Height & 672 & 672 & 672 \\
	Max batches & 60000 & 20000 & 20000 \\
	Learning rate & 0.001/0.0001 & 0.001/0.0001 & 0.001/0.0001\\ \bottomrule
	\end{tabular}}
	\label{table:3}
\end{table}

The training parameters are set for a balance of training performance and capability of the hardware. A high accuracy requires a high batch as 128, and the limitation of hardware sets the subdivisions and training size as 32 and 672×672, respectively. Max-batches are set to 60000 in the initial training and 20000 in the enhanced training. The minimum loss is found near 48000 batches in the initial training and 12000 batches in the enhanced training. The learning rate is set to 0.001 at the beginning of training, and a step decay policy is applied to decrease the learning rate from 0.001 to 0.0001 after 40000 batches in initial training and 8000 batches in enhanced training.

\subsection{Results of Vehicle Detections}

Vehicles are detected on the test videos using the calibrated model. When evaluating the integrity of the trajectory, we sample of 500 frames randomly, for counting and manually labeling the bounding boxes. In order to evaluate the performance of our training strategy, we calculate the results of each step in the training procedure, i.e. basic training, enhanced training, and duplication reduction. The vehicle detection performance is shown in Table\ref{table:4}.

\begin{table}
	\tbl{YOLOv4 Detection Results}
	{\begin{tabular}{l l l l l l l} \toprule
		\multicolumn{1}{c}{Performance} &
		\multicolumn{3}{c}{Test video \#1} &
		\multicolumn{3}{c}{Test video \#2} \\
		~ & Bsc train & Enh train & Dup reduc & Bsc train & Enh training & Dup reduc \\ \midrule
		Ground truth & \multicolumn{3}{c}{50998} & \multicolumn{3}{c}{125993} \\
		True positive & 46520 & 47698 &47535 & 105091 & 120046 & 119681 \\
		False negative & 6385 & 2954 & 3100 & 20902 & 5947 & 6312 \\
		False positive & 1132 & 6119 & 551 & 7257 & 24934 & 1424 \\
		Recall & 91.22\% & 93.53\%  & 93.21\%  & 83.41\%  & 95.28\%  & 94.99\%  \\
		~ &  (sd=0.0017) & (sd=0.0006) & (sd=0.0012) & (sd=0.0027) & (sd=0.0011) & (sd=0.0012) \\
		Precision & 97.63\% & 88.63\% & 98.86\% & 93.54\% & 82.80\% & 98.83\% \\ \midrule
		Average IoU &  \multicolumn{3}{c}{0.8377} & \multicolumn{3}{c}{0.8135}\\
		\bottomrule
	\end{tabular}}
	\label{table:4}
\end{table}

It shows that the YOLOv4 detection model after parameter selection and enhancement performs stable and shows the comprehensive result in UAV videos. Notably, the Recall in test video \#1 is 91.22\%, which is higher than that in test video \#2 (83.41\%). The main reason for the difference is that compared with video \#1 which is under the free-flow condition, video \#2 is under the congested condition and possesses larger vehicle amounts and smaller spacing which lowers the confidence score in the detection section. And the average IoU is only calculated by the true positive detection boxes. The two test videos are at a similar and good level although a vehicle only covers about 1500 pixels. We find that with basic training models, some dark vehicles and large vehicles are not accurately detected (Fig.~\ref{fig:3}).

Table~\ref{table:4} shows that the Recall in both videos increases apparently to 93.53\% and 95.28\% after the enhanced training. The main reason is that the separate neural network models successfully detect more dark and large vehicles. However, the Precision shows a significant decrease resulting from the overlap of vehicle detection boxes from different models (Fig.~\ref{fig:3}). 

After the duplication reduction, as shown in Table ~\ref{table:4}, the Precision successfully increases from 88.63\% to 98.86\% in test video \#1 and 82.80\% to 98.83\% in test video \#2, respectively. It means the false vehicle detections are successfully removed from the data. The Recall maintains at a high level, indicating that the duplication reduction does not wrongly delete true detections. The final Recall is 93.21\% in test video \#1 and 94.99\% in test video \#2, respectively. It indicates that the detection method of our framework works well in both congested and free-flow traffic conditions.

In the confusion matrix(Table~\ref{table:2}), the two types of errors are false negative and false positive. False negative indicates the missed detections of true vehicles. It usually happens when the lost vehicle has a special shape, color or vehicle type, vehicle type, or the vehicle is obscured by road decorations such as signs. This problem may be reduced when more training samples are equipped, and the number of samples of each type is the same. False positive indicates fake vehicle detections, are caused by the mix-up detection of the road surface or road markings with vehicles because of confusing features.

\begin{figure}
	\centering
	\includegraphics[width=0.9\linewidth]{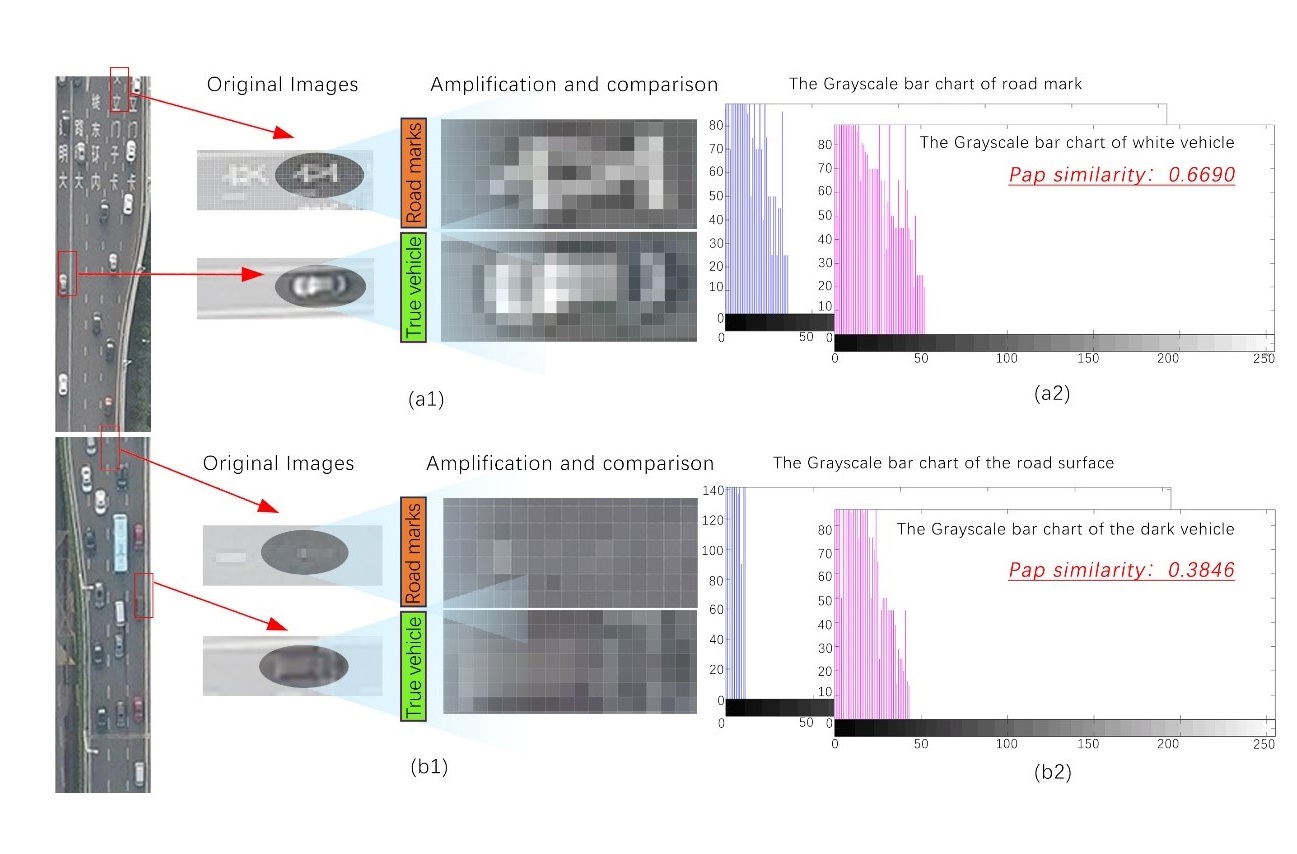}
	\caption{Two typical false examples detection errors.}
	\label{fig:9}
\end{figure}

Two typical examples are shown in Figure~\ref{fig:9}. In Figure~\ref{fig:9}(a1), the road marking is very similar to a white vehicle so that the detector makes a false positive detection. The pap similarity of the color histograms is 0.6690 in Figure~\ref{fig:9} (a2), which shows that the two pictures are very similar. In Figure~\ref{fig:9} (b1), the vehicle contour is not clear so that the detector did not recognize it and made a false negative mistake. Their color histograms possess a pap similarity of 0.3846 in Figure~\ref{fig:9} (b2).

% TODO: \usepackage{graphicx} required

\subsection{Results of Trajectory Construction}

Through trajectory construction, raw vehicle trajectories are extracted from two test videos. The results are shown in Table~\ref{table:5}. The research team manually counts the true count of trajectories, which is 500 and 541 in the two videos. After the trajectory construction, 484 and 497 trajectories are correctly obtained, while 4 and 43 trajectories are false positive ones. The improvement by integrity enhancement on trajectory count is evaluated in Table~\ref{table:5} (Trajectory Identification is simplified to T Identification and Integrity enhancement is simplified to I Enhancement). Results show that Recall increases from 92.20\% to 93.00\% and 83.92\% to 86.69\% in test video \#1 and test video \#2, respectively. Mainly, fifteen trajectories are put into integrity enhancement in test video \#2 while only four trajectories are enhanced in test video \#1.

This result shows a good performance in trajectory identification steps based on traffic flow theory. The majority of trajectories have been matched and constructed correctly. Besides, a large number of false positive and false negative in detection steps have been eliminated in the trajectory construction step.

\begin{table}
	\tbl{Trajectory Construction Performances}
	{\begin{tabular}{l l l l l} \toprule
		\multicolumn{1}{c}{Performances} &
		\multicolumn{2}{c}{Test Video \#1} &
		\multicolumn{2}{c}{Test Video \#2}\\
		~ & \multicolumn{1}{c}{T Identification} &
		\multicolumn{1}{c}{I Enhancement} &
		\multicolumn{1}{c}{T Identification} &
		\multicolumn{1}{c}{I Enhancement} \\ \midrule
		Ground Truth &  \multicolumn{2}{c}{500} & \multicolumn{2}{c}{541}\\
		True positive & 461 & 465 & 454 & 469  \\
		False negative & 39 & 35 & 87 & 72 \\
		False positive & 4 & 4 & 43 & 43 \\
		Recall & 92.20\% & 93.00\% & 83.92\% & 86.69\% \\
		Precision & 99.13\% & 99.15\% & 91.35\% & 91.60\% \\
		\bottomrule
	\end{tabular}}
	\label{table:5}
\end{table}

In the two videos, 107 trajectories were missing or incorrectly constructed. We carefully checked these errors and found that most of the missing trajectories are caused by a long period of detection failure. The continuum of detection boxes in the trajectories is insufficient, so our methods are not able to form the complete trajectories. Only a few missing trajectories are caused by the detection box in one trajectory identifying wrongly to the trajectory of another vehicle. The mismatch occurs when the to-be-identified detection box is lost in the next several frames and too much position derivation is needed. At this time, the interference of adjacent vehicles will increase, making it easier to misidentify. 

\subsection{Results of Trajectory Denoising}

It is very difficult, or impossible, to obtain the ground truth of the vehicle trajectories for the validation of the accuracy of constructed trajectories. As a result, in our study, we estimate the validity of the denoising procedure by checking the reasonableness of the distribution range of parameters such as vehicle speed, acceleration, space headway, time headway, and gap.

\begin{figure}
	\centering
	\includegraphics[width=0.8\linewidth]{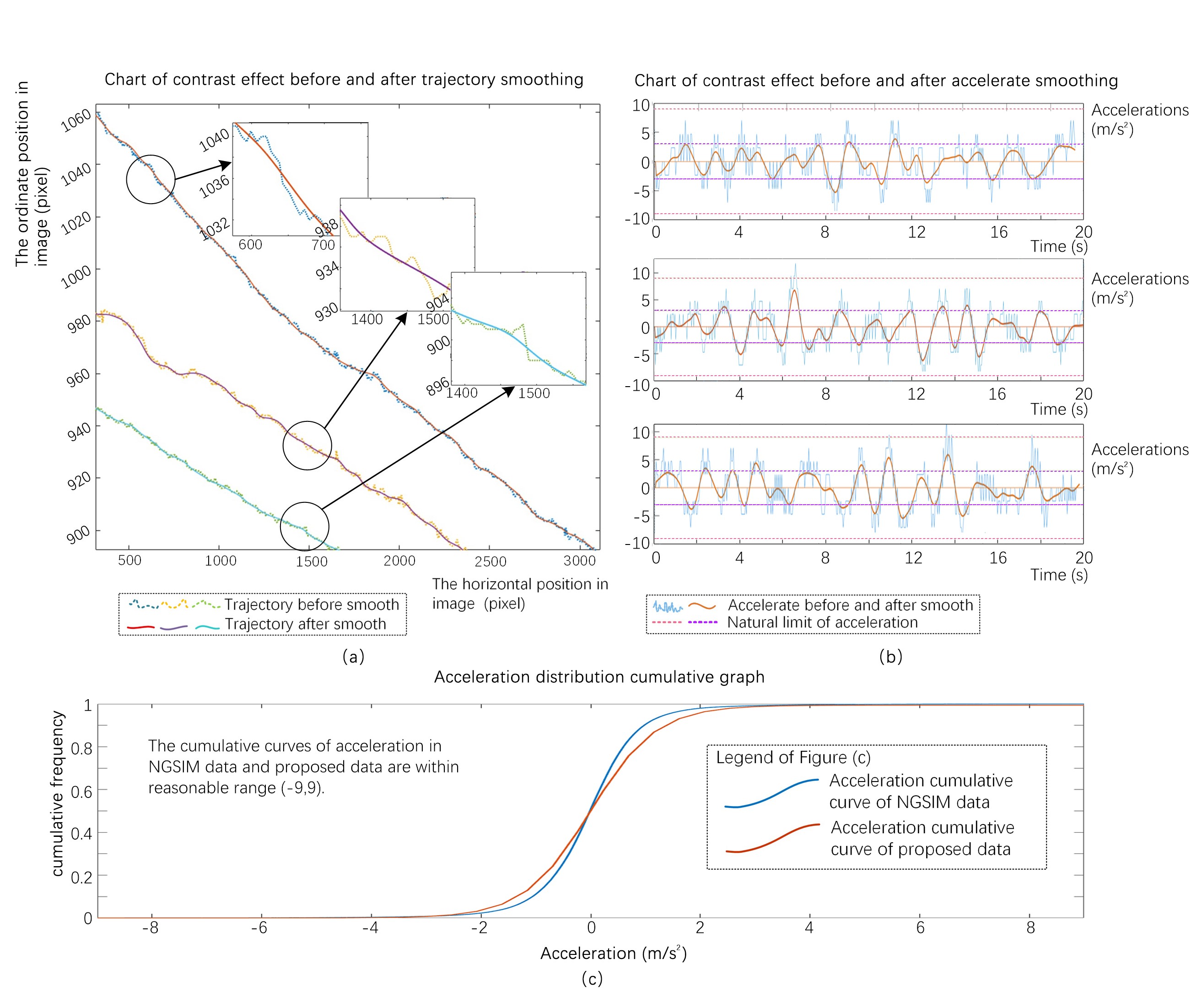}
	\caption{The vehicle trajectory denoising result.}
	\label{fig:10}
\end{figure}

A typical vehicle trajectory denoising result using EEMD is shown in Figure~\ref{fig:10}. The contribution of smoothing out the noise can be seen in Figure~\ref{fig:10}. The raw trajectory data has large fluctuation, while the denoised data is much smoother. The acceleration change resulting from the denoising process is shown in Figure~\ref{fig:10} (b). According to \citet{Montanino:2013}, the physical limitations of vehicle acceleration and deceleration rate are $9 m/s^2$ and $-9 m/s^2$, respectively. And as our know, the comfortable acceleration are within $\pm 3m/s^2$. Our denoised acceleration data is all within a reasonable range (between the pink dotted lines), and most of them are distribute in the range of comfortable accelerations (between the purple dotted lines). The cumulative percentage curve of the acceleration of all vehicles from our data and from Montanio’s method is shown in Figure~\ref{fig:10} (c). It can be clearly seen that the distributions of the acceleration of two trajectory data after denoising are consistent.

% TODO: \usepackage{graphicx} required

\begin{figure}
	\centering
	\includegraphics[width=0.8\linewidth]{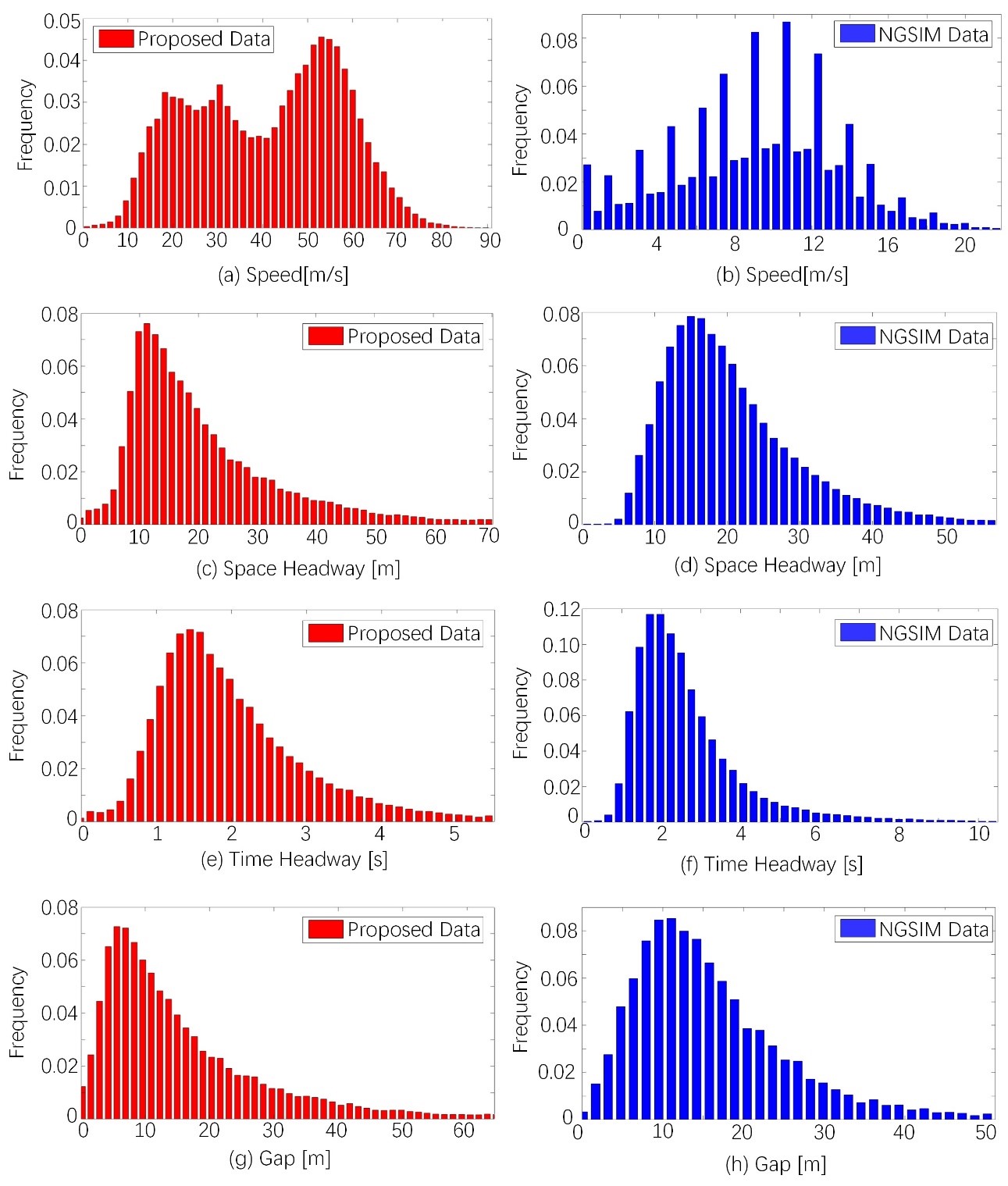}
	\caption{Comparison of key traffic parameters with NGSIM.}
	\label{fig:11}
\end{figure}

To further validate the reasonableness of our data, we make a comparison of the value of key parameters (speed, space headway, time headway, and gap) with the NGSIM dataset from the southbound traffic on US-101 in Los Angeles, CA, recorded from 14:50-15:05, on June 15, 2005. The visualization of the comparison is shown in Figure~\ref{fig:11}. As seen in Figure~\ref{fig:11} (c-d), the space headway with the highest frequency in our data (the red bar) is around 12m, while that in NGSIM data (the blue bar) is around 40m. Similarly, in Figure~\ref{fig:11}(e-h), the time headway with the highest frequency is around 1.6s in both data, and the gap with the highest frequency is 8m and 40m in our data and NGSIM data, respectively. The difference in the value of the gap and the space headway maybe because our test videos partially suffer from congestions. For the same reason, the speed of our data (Figure~\ref{fig:11} (c)) has two peaks around $30km/h$ and $55km/h$. The speed of NGSIM data has one peak around 40km because the data were collected from the free-flow traffic state. 

% TODO: \usepackage{graphicx} required

\subsection{Evaluation of Processing Speed}

The processing speed of our framework for trajectory extraction is calculated. The majority of time is spent on the offline training of the YOLOv4 algorithm. The training time is about 70h, which is similar to the previous study \citep{Benjdira:2019}. After the training, the online processing speed is very fast. The processing time in each step of our framework is shown in Table ~\ref{table:6}. The detection step only costs 0.1 h, suggesting the remarkable advantage of applying YOLOv4. The trajectory construction costs 3.8 h (26s/track) and 4.0 h (27s/track), including the trajectory identification, integrity enhancement and coordinate transformation. The denoising step is fast, which costs only 0.3 h (2s/track). 

In summary, the total processing time is 4.2 h and 4.3 h on the test video \#1 and \#2. The slightly longer processing time in video \#2 is caused by the fact that there are more vehicles in congested traffic conditions. The average online processing speed is 29s/track and 30s/track, or in other words 0.6 s/frame and 0.7 s/frame, on the two test videos.

\begin{table}
	\tbl{Processing Time of Our Framework}
	{\begin{tabular}{l l l l l} \toprule
	Section &
	\multicolumn{2}{c}{Processing time} &
	\multicolumn{2}{c}{Processing speed}\\
	~ & \multicolumn{1}{c}{Test video \#1} &
	\multicolumn{1}{c}{Test video \#2} &
	\multicolumn{1}{c}{Test video \#1} &
	\multicolumn{1}{c}{Test video \#2} \\ \midrule
	Training &  \multicolumn{2}{c}{70h} & \multicolumn{2}{c}{13s/vehicle}\\
	Image stabilization & \multicolumn{2}{c}{40 minutes} & \multicolumn{2}{c}{0.3s/frame} \\
	Detection & 0.1h & 0.1h & $6*10^{-3}s/vehicle$ & $3*10^{-3}s/vehicle$ \\
	T Identification & 1.3h & 1.5h & 9s/track & 10s/track \\
	I Enhancement & 0.5h & 0.5h & 3s/track & 3s/track \\
	C transformation & 2h & 2h & 14s/track & 14s/track \\
	Denosing & 0.3h & 0.3h & 2s/track & 2s/track \\ \midrule
	whole framework & 4.2h &4.4h & 29s/track & 30s/track \\ 
	(without training) &  & & &  \\
	\bottomrule
	\end{tabular}}
	\label{table:6}
\end{table}

\section{CONCLUSIONS AND DISCUSSION}

A novel framework for massive vehicle trajectory matching and construction from aerial videos is proposed in this research. This framework is based on the vehicle dynamic constraints derived from Newell’s traffic flow theory. In the framework, an integrated high-precision vehicle detection by YOLOv4 is arranged and strengthened. Unique trajectory decision-making and reconstruction method is applied to detected bounding boxes frame-by-frame. This step is equipped with trajectory identification, integrity enhancement and coordinate transformation from image coordinates to the Frenet coordinates. Then the raw trajectory obtained from the construction section will be denoised by EEMD. Our framework is tested on two aerial videos taken by a UAV on a city expressway covering congested and free-flow traffic conditions. The extracted vehicle trajectories are compared with manual counts and labeled bounding boxes.

The experimental results showed that the average Recall of vehicle detection count in the two videos are 93.21\% and 94.99\%, respectively. And the average Precision of vehicle detection are 98.86\% and 98.83\%. respectively. IoU of the vehicle detection is 0.8377 and 0.8135 respectively. The Recall of trajectory construction count was 93.00\% and 86.69\% while the Precision of trajectory construction is 99.15\% and 91.60\% in the two videos. The denoising algorithm performed effectively in eliminating outliners and keeping the traffic parameters within a reasonable range. The total processing time of our framework was 4.2 h and 4.4 h on test videos \#1 and \#2. 

\begin{figure}
	\centering
	\includegraphics[width=1\linewidth]{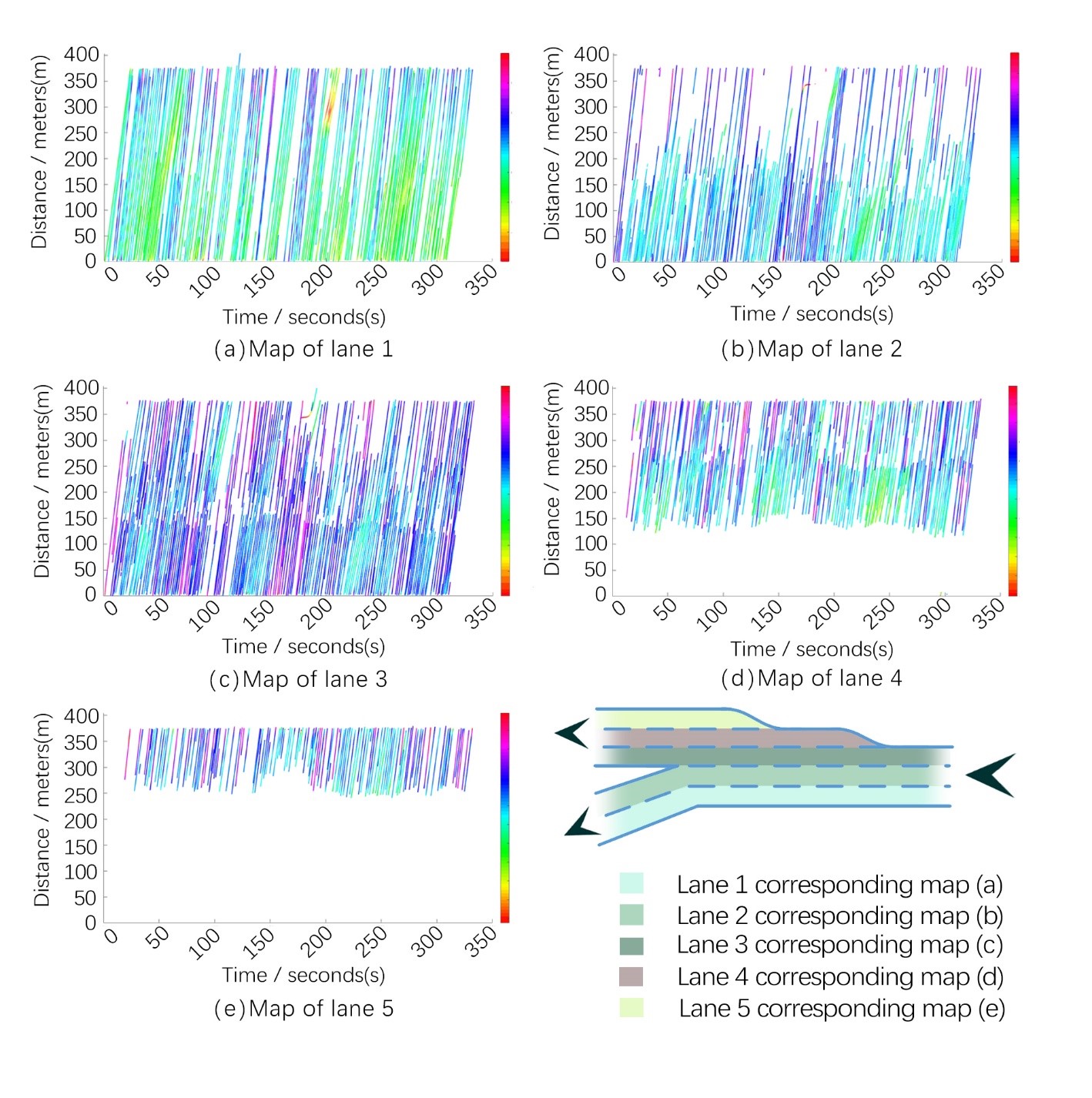}
	\caption{Space-time trajectory maps (test video \#1).}
	\label{fig:12}
\end{figure}

\begin{figure}
	\centering
	\includegraphics[width=1\linewidth]{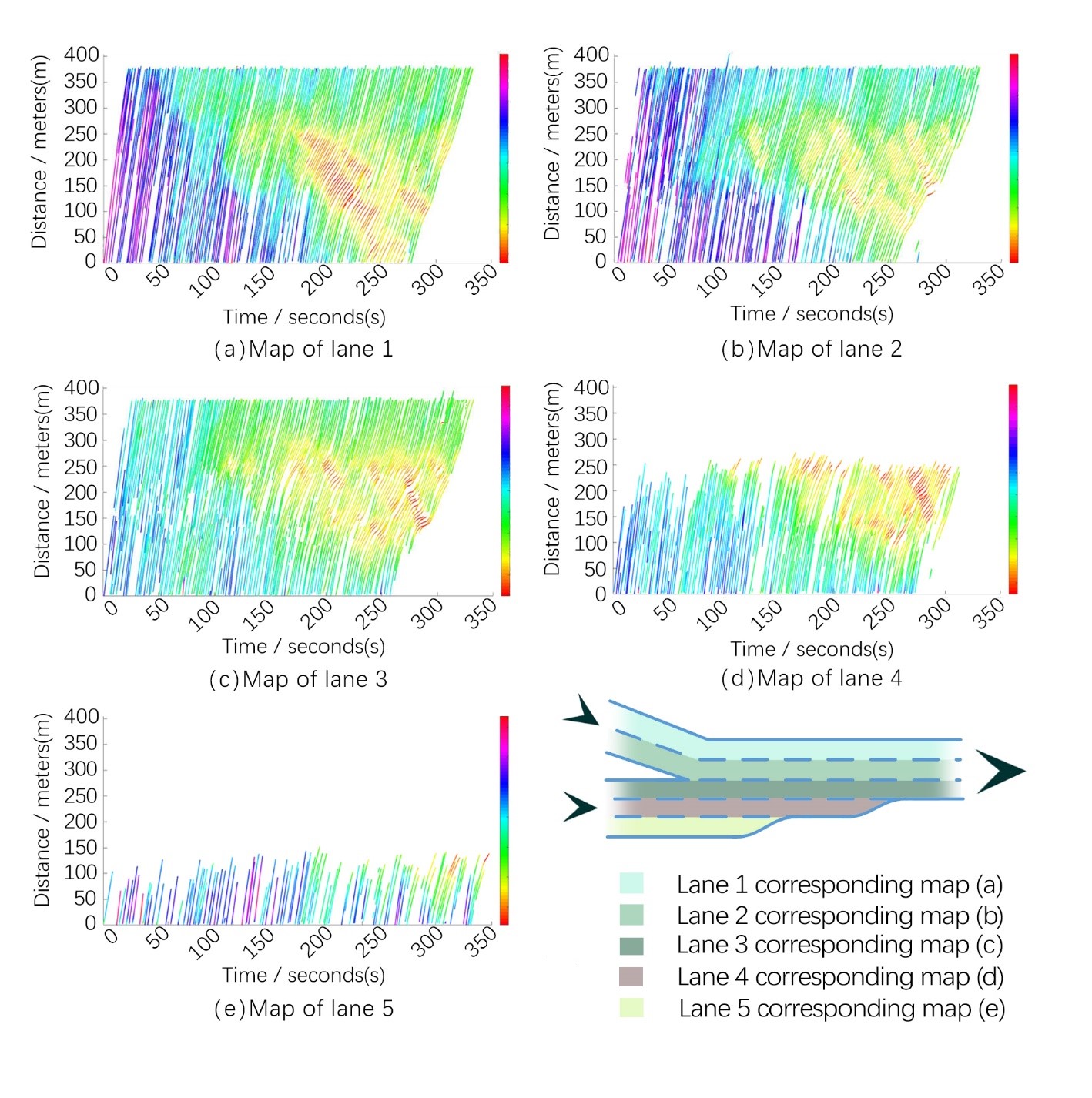}
	\caption{Space-time trajectory maps (test video \#2).}
	\label{fig:13}
\end{figure}

Our methodology can help enrich the trajectory data for traffic flow studies. Based on our video data, we construct the space-time trajectory map of each lane, as shown in Figure~\ref{fig:12}, ~\ref{fig:13}. It can be observed that the traffic is free-flowing in video \#1 with a relatively constant speed on each lane. While in video \#2, the formation of congestion as well as kinematic waves is evident. The critical car-following and lane-changing parameters can be estimated from the trajectory map for more comprehensive traffic flow modeling and analysis. 
% TODO: \usepackage{graphicx} required

% TODO: \usepackage{graphicx} required

%%%
%\section{ACKNOWLEDGMENTS}
%This research was sponsored by the National Natural Science Foundation of China (71871057). The authors would like to thank Mr. Guoqing Peng and Mr. Daan Zhu for their efforts on the aerial video collection. And we are thankful for Mr. Ou Zheng’s generous and useful advisements on the experiment. Thanks to Mr. Siqi Feng and Mr. Yuqing Cheng for their effective suggestions on typesetting. Other crew members should be thanked for the manual labeling work. 

%\section{DATA AVAILABILITY STATEMENT}
%The video processed by this research was taken by UAV (model: DJI Mavic professional) at 280m and 310m above Nanjing expressway. We have created an open-access website platform for sharing the videos as well as the trajectory data we have extracted. These data can be accessed at \href{http://seutraffic.com/}{http://seutraffic.com/}. More published trajectory data can also be obtained from this website.

\bibliography{interactapasample}

\end{document}